\documentclass[aps,pra,twocolumn,showpacs,superscriptaddress]{revtex4-1}

\usepackage{amsfonts,mathrsfs,amsmath,amsthm,amssymb,bbold}                   
\usepackage{epsfig}
\usepackage{bm}
\usepackage{xfrac}
\usepackage{tabulary}
\usepackage{color}
\usepackage[dvipsnames]{xcolor}
\usepackage{graphicx}

\usepackage[colorlinks=true,linkcolor=blue,citecolor=blue,urlcolor=magenta]{hyperref}

\begin{document}  
\title {\bf Harnessing graph state resources 
for robust quantum magnetometry under noise}

\author{Phu Trong Nguyen}
%\thanks{These authors contributed equally to this work.}
\affiliation{Department of Advanced Material Science and Nanotechnology, 
University of Science and Technology of Hanoi, Vietnam Academy of Science and Technology, Hanoi, Vietnam}

\author{Trung Kien Le}
%\thanks{These authors contributed equally to this work.}
\thanks{Current address: Department of Applied Physics, Stanford University, Stanford, USA}
\affiliation{Department of Physics, University of California, Santa Barbara, 
Santa Barbara, USA}

\author{Hung Q. Nguyen}
\affiliation{Nano and Energy Center, University of Science, 
Vietnam National University, Hanoi, 120401, Vietnam}

\author{ Le Bin Ho}
\thanks{Electronic address: binho@fris.tohoku.ac.jp}
\affiliation{Frontier Research Institute for Interdisciplinary Sciences, 
Tohoku University, Sendai 980-8578, Japan}
\affiliation{Department of Applied Physics, Graduate School of Engineering, 
Tohoku University, Sendai 980-8579, Japan}

\date{\today}

\begin{abstract}
Precise measurement of magnetic fields is essential for various applications, such as fundamental physics, space exploration, and biophysics. Although recent progress in quantum engineering has assisted in creating advanced quantum magnetometers, there are still ongoing challenges in improving their efficiency and noise resistance. This study focuses on using symmetric graph state resources for quantum magnetometry to enhance measurement precision by analyzing the estimation theory under time-homogeneous and time-inhomogeneous noise models. The results show a significant improvement in estimating both single and multiple Larmor frequencies. In single Larmor frequency estimation, the quantum Fisher information spans a spectrum from the standard quantum limit to the Heisenberg limit within a periodic range of the Larmor frequency, and in the case of multiple Larmor frequencies, it can exceed the standard quantum limit for both noisy cases. This study highlights the potential of graph state-based methods for improving magnetic field measurements under noisy environments. 
\end{abstract}

%
%\pacs{03.65.Ta, 06.20.-f, 42.50.Lc}
% Quantum metrology, Graph states,  Quantum measurements,

\maketitle

\section{Introduction}\label{seci} 
Quantum sensing utilizes quantum resources like 
non-classical states, 
entanglement, and squeezing to 
improve sensor capabilities beyond classical approaches
\cite{RevModPhys.89.035002}. 
Recent advances in quantum resource theory 
have been made in quantum-enhanced sensing 
using non-classical states 
%of highly magnetic atoms 
\cite{Chalopin2018, RevModPhys.90.035005, HO2019153}, 
entangled cluster and graph states
\cite{Friis_2017,PhysRevLett.124.110502,PhysRevA.102.052601,Le2023}, 
many-body nonlocality and multiqubit systems
\cite{PhysRevLett.126.210506,PhysRevLett.130.170801,PhysRevResearch.5.043097}, 
and squeezed resources \cite{Zhang_2014,Maccone2020squeezingmetrology,Gessner2020,PhysRevLett.131.133602}. 
Furthermore, various techniques 
like machine learning algorithms 
\cite{PhysRevLett.123.230502,Jung2021,Costa2021,10.1117/1.AP.5.1.016005,rinaldi2023parameter}, 
quantum error correction methods \cite{PhysRevLett.112.150802,Shettell_2021,PhysRevLett.129.250503,PhysRevLett.128.140503}, 
network sensing \cite{PhysRevLett.120.080501,Rahim2023}, 
and hybrid algorithms \cite{Koczor_2020,Yang2021,PhysRevX.11.041045,Meyer2021,Marciniak2022,Le2023,PRXQuantum.4.020333}, 
are being explored for enhancing noise resilience 
and extracting insights from quantum sensing. 
%These advancements can revolutionize quantum sensing 
%applications and open new avenues for high-precision 
%measurements across diverse fields.

%So far, there have been various developments in quantum resources and measurement technologies to improve precision. These approaches include adaptive measurements, error correction, and optimal quantum control.

In quantum magnetometry, 
the precise measurement of magnetic fields is 
crucial in various subjects like 
fundamental physics research, space exploration, 
material science, geophysics, and medical biophysics. 
Recent advances in quantum engineering 
have led to the development of various 
quantum magnetometers, such as 
superconducting quantum interference device (SQUID) \cite{Portoles2022},
diamond-based magnetometer \cite{Gulka2021, Carmiggelt2023}, 
single-spin quantum magnetometer \cite{Huxter2022}, 
submicron-scale NMR spectroscopy \cite{Sahin2022}, 
cold atom magnetometer \cite{10.1116/1.5120348},
and 2D hexagonal boron nitride magnetic sensor 
\cite{PhysRevApplied.18.L061002}.
These innovations find applications in highly sensitive 
and broadband magnetic field measurements 
\cite{Gulka2021, Carmiggelt2023}, 
scanning gradiometry \cite{Huxter2022}, 
low magnetic fields \cite{Sahin2022}, 
navigation \cite{10.1116/1.5120348},
and magnetic field imaging \cite{PhysRevApplied.18.L061002}.

Improving the sensitivity of magnetometers 
is essential for different applications. 
However, the present methods are insufficient 
due to high quantum resource efficiency 
and noise resilience demands. Therefore, 
there is an urgent requirement for 
a novel resource that can enable 
the full potential of quantum magnetometry 
while being practical for experimental implementation.

Among various candidates, graph states have emerged 
as a promising avenue in the quest for 
quantum-enhanced magnetometry. 
Graph states are particular types of entangled states 
that can be represented by a graph, 
where the vertices represent qubits, 
and the edges represent entangling gates 
between the qubits \cite{Hein:2006,PhysRevA.69.062311}. 
Due to their multipartite entanglement, 
they have demonstrated great potential 
in quantum computation \cite{Bell2014,PhysRevA.65.012308}, 
communication \cite{Bell2014,Bell2014_s}, 
and metrology \cite{PhysRevLett.124.110502,PhysRevA.102.052601,Le2023}. 
Within the context of quantum magnetometry, 
harnessing the capabilities of graph states 
introduces a novel dimension to the quest 
for precision and robustness in the presence of noise.

This work explores symmetric graph state resources 
for robust quantum magnetometry under time-homogeneous and time-inhomogeneous noises \cite{PhysRevA.102.022602}. Our approach begins 
by modeling an ensemble of $N$ spin-1/2 particles 
as a sensor probe for measuring Larmor frequencies 
of an external magnetic field. Initially, 
the probe state is set up in a star-graph configuration, 
where one vertex (spin particle) is connected to the remaining 
$N-1$ vertices through CZ gates. 
We study the influence of noise 
in the model by analyzing 
the measurement precision  
from the perspective of estimation theory 
and quantum Fisher information.

For uncorrelated probes, 
the variance of estimating a single phase $\phi$ 
follows $\Delta^2\phi = \mathcal{O}(N^{-1})$, 
commonly referred to as the standard quantum limit (SQL), 
whereas for entangled probes, 
it is possible to reach the Heisenberg limit (HL), 
where $\Delta^2\phi = \mathcal{O}(N^{-2})$ 
\cite{PhysRevLett.102.100401,doi:10.1126/science.1104149,PhysRevLett.96.010401}. 
However, under 
time-homogeneous noise, 
entangled sensors cannot surpass the SQL 
\cite{PhysRevLett.79.3865,PhysRevLett.116.120801}. 
In the presence of 
time-inhomogeneous noise, 
the variance can reach $\Delta^2\phi = \mathcal{O}(N^{-1.5})$ 
\cite{PhysRevA.84.012103,PhysRevLett.109.233601,PhysRevLett.115.170801}, 
and similar results have been observed in the context of multiphase sensing
\cite{PhysRevA.102.022602,Le2023}.

In our investigation for single Larmor frequency estimation, 
we observe a transition from SQL to HL behavior 
for a periodic range of the Larmor frequency.
%represented as $\Delta^2\phi = \mathcal{O}(N^{-L})$ with $L \approx 2$. 
For multiple  Larmor frequencies, we find that the variance 
$\Delta^2\bm\phi$ can %reach $\mathcal{O}(N^{-1.5})$ 
beat the SQL
for both 
time-homogeneous and time-inhomogeneous noise sources. 
This marks the initial instance where 
we observe surpassing the SQL under time-homogeneous noise.
Our analysis of quantum magnetometry in these noise factors sheds light on the resilience 
and potential of graph state-based approaches 
for conducting highly precise magnetic field measurements 
in challenging and real-world conditions.
\section{Results}\label{secii}
\subsection{Measurement model and its initialization}
Let us consider the measurement of an external magnetic field 
by employing a spin-1/2 system comprising $N$ 
particles as the probing mechanism. 
Each particle interacts with the field 
and provide information about the field strengths. 
The coupling Hamiltonian is given by
%The magnetic field couples with 
%all spins in the probe via the Hamiltonian 
\cite{https://doi.org/10.1002/mrc.1092}
\begin{align}\label{eq:H1}
    \mathcal H = -\sum_{k = 1}^N 
    \Big(\bm\mu^{(k)}\cdot\bm B\Big),
    %= \sum_{k = 1}^N 
    %\Big(\frac{1}{2}
    %|\gamma_e|\bm\sigma^{(k)}\cdot\bm B\Big)
\end{align}
where $\bm\mu^{(k)} = \frac{1}{2}\gamma_e\bm \sigma^{(k)}$ 
represents the magnetic moment of 
the $k^{\rm th}$ spin. $\gamma_e$ 
denotes the gyromagnetic ratio, 
and $\bm \sigma^{(k)} =
\big(\sigma_x^{(k)},\sigma_y^{(k)},\sigma_z^{(k)}\big)$ 
refers to the Pauli matrices. 
Here, $\bm B = (B_x, B_y, B_z)$ 
signifies the external magnetic field.
We define $\phi_j = |\gamma_e|B_j$ 
for all $j \in {x, y, z}$ 
as the Larmor frequency
\cite{https://doi.org/10.1002/mrc.1092}, 
and $J_j = \frac{1}{2}\sum_k \sigma_j^{(k)}$ 
as the angular momentum. With this, 
the Hamiltonian recasts as
\begin{align}\label{eq:Ht}
    \mathcal H 
    = \bm\phi\cdot\bm J,
\end{align}
where $\bm\phi = (\phi_x, \phi_y, \phi_z)$ 
represents the set of Larmor frequencies 
requiring estimation, 
and $\bm J = (J_x, J_y, J_z)$ 
are three components 
of the collective angular momentum.
Refer to %App.~\ref{appA} 
the Methods section
for a detailed model and its quantum circuit.

The probe is initialized as a graph state, 
which typically consists of 
a collection of vertices denoted by 
$V$ and edges represented by $E$ as
%A conventional graph state is formed by a collection 
%of vertices $V$ and edges $E$ as 
\begin{align}\label{graph_state}
G(V,E) =  \prod_{{i,j} \in E} {\rm CZ}_{ij}  |+\rangle ^{\otimes V},
\end{align}
where ${\rm CZ}_{ij}$  represents the controlled-Z gate  
connecting the $i^{\rm th}$ and $j^{\rm th}$ spins, 
and $|+\rangle$ is an element in 
the basis of Pauli $\sigma_x$. 
Graph states serve as valuable 
assets in quantum metrology
\cite{PhysRevLett.124.110502,Le2023}, 
as demonstrated by their application 
in achieving Heisenberg scaling, 
as observed with star configurations 
where the quantum Fisher information (QFI) gives $(N-1)^2 +1$, 
and with local Clifford (LC) operations where 
the QFI gives $N^2$ \cite{PhysRevLett.124.110502}. 
Hereafter, we examine the impact of 
graph-state resources on 
quantum-enhanced magnetometry 
within a noisy environment.

%\subsection{Estimation theory and quantum Fisher information}
%In the estimation theory, the precision is 
%characterized by a covariance matrix 
\begin{figure}[t!]
    \centering
    \includegraphics[width=8.6cm]{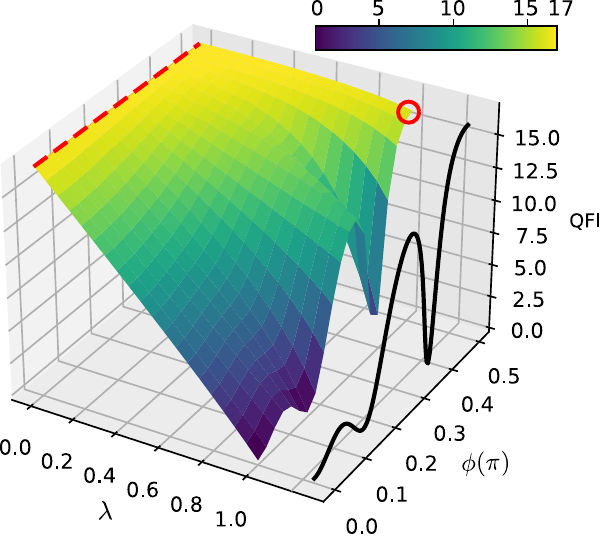}
    \caption{
    \textbf{Quantum Fisher information in a single phase estimation.}
    Plot of QFI vs the noise probability $\lambda$
    and Larmor frequency $\phi$ for $N = 5$ and $t = 1$.
    For $\lambda = 0$, the QFI reaches $Q = (N-1)^2 + 1\ \forall \phi$ 
    (dashed red line). Increasing $\lambda$, the QFI gradually reduces
    and reaches the minimum at $\lambda = 1$. 
    Remarkably, this minimum is non-zero for a non-zero $\phi$ 
    as illustrated by the soiled black curve. 
    For $\phi = \pi/2$, it is given by $(N-1)^2$ (red circle).
    }
    \label{fig:1}
\end{figure}

\subsection{Single phase estimation}
We examine the estimation of 
a single Larmor frequency 
denoted as $\bm \phi = (\phi, 0, 0)$. 
The coupling Hamiltonian is 
$\mathcal H = \phi J_x$, 
and the corresponding unitary 
operator is expressed as 
\begin{align}\label{eq:uni} 
\mathcal U(\phi) = \exp\big(-it\phi J_x\big). 
\end{align}
%where $\phi$ is the unknown phase
%and $\bm J_x = \frac{1}{2}\sum_{k = 1}^N\sigma_x^{(k)}$ 
%with $\sigma_x^{(k)}$ is the 
%Pauli matrix acting on the $k^{\rm th}$ 
%qubit. 
The initial probe state is prepared in 
a graph configuration $\rho_0 = 
|G\rangle\langle G|$.
After the interaction, 
it evolves to $\rho(\phi) = 
\mathcal U(\phi)\rho_0\mathcal U^\dagger(\phi)$.
During the magnetic field coupling, 
the probe interacts with its surroundings 
and decoheres. 
Our analysis focuses on dephasing noise 
as a type of phase decoherence that leads to 
the evolution of the state
\begin{align}\label{eq:rho1}
    \rho({\phi,\gamma}) = 
    \Big[\prod_{k = 1}^N 
    e^{\gamma t\mathcal{L}^{(k)}}\Big]
    \rho(\phi),
\end{align}
where $\gamma$ is the dephasing rate.
Dephasing is often referred to as the spin-spin relaxation process
\cite{NielsenChuang2010},
which affects the relative phase in the probe’s basis 
and can be represented by the Pauli operator $\sigma_z$.
By employing the Kraus operators 
to account for dephasing noise as
$\mathcal K_0 = {\rm diag}(1, \sqrt{1-\lambda}),
\mathcal K_1 = {\rm diag}(0, \sqrt{\lambda})$,
we obtain
\begin{align}\label{eq:kraus}
e^{\gamma t\mathcal{L}^{(k)}}\rho(\phi)
= \mathcal K_0^{(k)}\rho(\phi)\big[\mathcal K_0^{(k)}\big]^\dagger+
\mathcal K_1^{(k)}\rho(\phi)\big[\mathcal K_1^{(k)}\big]^\dagger,
\end{align}
where $\lambda = 1 - e^{-\gamma t} \in[0,1]$ 
is the dephasing probability. 
%Here, we used $\rho$ to denote $\rho(\phi)$.
For other noisy scenarios, please see %App.~\ref{appD}.
the Methods section.

The final state $\rho({\phi,\gamma})$ 
contains detailed information about 
the unknown Larmor frequency $\phi$. 
To evaluate the precision of the estimation, 
we examine the QFI $Q$.
By decomposing $\rho({\phi,\gamma}) 
= \sum_k \ell_k |\ell_k\rangle\langle\ell_k|$, 
the QFI yields \cite{doi:10.1142/S0219749909004839}
\begin{align}\label{eq:qfi}
 Q = 2\sum_{i,j,\ell_i+\ell_j\ne 0} 
 \dfrac{\big|\langle\ell_i| 
 \partial_\phi\rho({\phi,\gamma}) 
 |\ell_j\rangle\big|^2}{\ell_i+\ell_j}. 
\end{align}

For numerical calculation,
let us fix the sensing time $t = 1$ in an arbitrary unit.
%which is suitable for spin systems. 
%By this choice, the dephasing rate $\gamma$
%and the Larmor frequency $\frac{\phi}{2\pi}$
%are in the unit of ms$^{-1}$ or kHz.
The results are presented in Fig.~\ref{fig:1}, 
focusing on the star-graph configuration 
and using $N = 5$ as an illustrative example. 
%Here, the time is fixed at $t = 1$. 
In the absence of noise, i.e., $\lambda  = 0$, 
the QFI yields \cite{PhysRevLett.124.110502}
\begin{align}\label{eq:Q1}
Q = 4\Big[\langle G| J_x^2 |G \rangle 
    - \big(\langle G| J_x |G \rangle\big)^2\Big]
= \big(N-1\big)^2+1,
\end{align}
which does not depend on $\phi$
as shown in the dashed red line. 
See %App.~\ref{appB} 
the Methods section
for detailed calculation. 

In the presence of noise, 
the QFI depends on both $\lambda$ and  $\phi$. 
As $\lambda$ increases, 
the QFI gradually decreases, 
reaching its minimum value at $\lambda = 1$. 
Interestingly, this minimum value remains 
nonzero for $\phi \ne 0$, 
which is demonstrated by the solid black curve. 
Specifically, for $\phi = \pi/2$, %$\rm {rad} \cdot \rm{ms}^{-1}$
%or $\frac{\phi}{2\pi} = 0.25$ kHz, 
the QFI is even by $(N-1)^2$ (red circle).
See %App.~\ref{appB} 
the Methods section
for detailed calculation. 

A specific case of star graph is a GHZ state
up to a local unitary (LU)
transformation \cite{PhysRevA.69.062311,PhysRevLett.91.107903}.
Let us consider the initial probe state to be a GHZ state
\begin{align}\label{eq:ghz}
|\psi\rangle_{\rm GHZ} 
= \dfrac{1}{\sqrt{2}} 
\Big(|\nu_{\max}\rangle 
+|\nu_{\min}\rangle \Big), 
\end{align} 
where $|\nu_{\max}\rangle$ 
and $|\nu_{\min}\rangle$ 
are eigenstates of $J_x$ 
corresponding to the maximum 
and minimum eigenvalues 
$\nu_{\max}$ and $\nu_{\min}$, respectively.
Particularly, this state can be 
prepared by applying a Hadamard gate 
to the first spin particle
of the star graph state in Eq.~\eqref{eq:wpa}.
The QFI gives %(see App.~\ref{appB})
(see the Methods section)
\begin{align}\label{eq:Qghz}
   Q_{\rm GHZ} = 4\Big[\langle \psi_{\rm GHZ}| J_x^2 |\psi_{\rm GHZ} \rangle 
    - \big(\langle \psi_{\rm GHZ}| J_x |\psi_{\rm GHZ} \rangle\big)^2\Big]
    = N^2.
\end{align}
Here, the QFI remains independent of $\phi$.
Upon closer examination, 
it becomes evident that the QFI attains 
the Heisenberg limit of $N^2$ 
in the absence of noise. 
%However, it diminishes to 
%the standard quantum limit 
%of $N$ under the influence of noise, 
%as detailed in App.~\ref{appB}. 
In the presence of dephasing noise, 
the QFI is invariant, 
preserving its $N^2$ 
as detailed in %App.~\ref{appB}.
the Methods section.
This result is trivial as noise primarily affects 
the phase or coherence of quantum states 
along the $z$-axis, 
while GHZ here points toward the $x$-axis.

Next, we examine the quantum Cram{\'e}r-Rao bound 
(QCRB) for various values of $N$. 
It is the ultimate bound that imposes the precision achievable in 
the estimation process, i.e, 
$M\cdot \Delta^2\phi \ge \mathsf{C_F} \ge \mathsf{C_Q}$,
where $\Delta^2\phi = \langle (\phi - \hat\phi)^2\rangle 
- \langle (\phi - \hat\phi)\rangle^2$ is the 
variance of $\phi$, 
which indicates the difference between the true value $\phi$ and 
its estimated counterpart $\hat\phi$,
$M$ is the repeated experiments. 
Here, $\mathsf{C_F}$ and $\mathsf{C_Q}$ 
are classical and quantum Cram{\'e}r-Rao bound, respectively. 
The QCRB is determined through the inversion 
of the QFI as %designated as 
\begin{align}
   \mathsf{C_Q} = Q^{-1},
\end{align}
which can be achieved in the single-phase estimation,
such as using a Bayesian estimator or neural network technique
(see \cite{rinaldi2023parameter} and Refs therein).

\begin{figure}[t!]
    \centering
    \includegraphics[width=8.6cm]{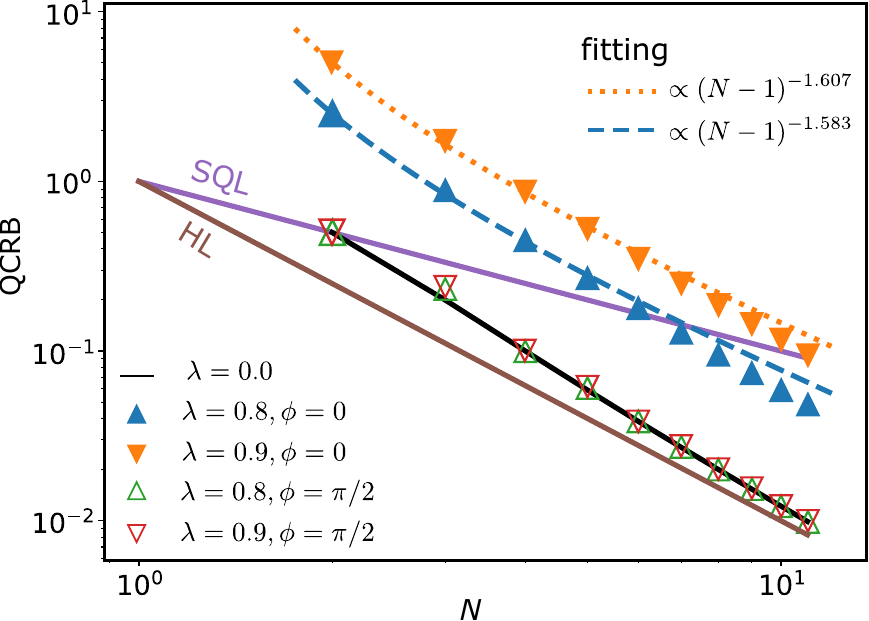}
    \caption{
    \textbf{Quantum Cram{\'e}r-Rao bound in a single phase estimation.}
    The plot of QCRB as a function of the number of spins $N$
    for various $\lambda$ and $\phi$. %The fitting curves for these cases 
    %indicate the super-Heisenberg scaling, i.e., $\propto N^{-L}, L > 2$.
    Additionally, SQL and HL are displayed for comparative purposes. 
    The plot is presented with the star-graph configuration.}
    \label{fig:2}
\end{figure}

The numerical results are showcased in Fig.~\ref{fig:2}.
For $\phi = 0$, the QCRB can beat the SQL event for large noise.
Here we illustrate for $\lambda = 0.8$ (\textcolor{blue}{\small$\blacktriangle$}) 
and 0.9 (\textcolor{orange}{\small$\blacktriangledown$}), and
the fitting curves are proportional to $(N-1)^{-1.583}$
(blue dashed line)
and $(N-1)^{-1.607}$ (orange dotted line), respectively.
Throughout the paper, we use
the fitting function as $f(N) = a(N-1)^b\ \forall a,b\in \mathcal R$,
which is inspired by the exact result when $\lambda = 0$.
For $\phi = \pi/2$, the analytical findings in Fig.~\ref{fig:1} 
suggest that the QCRB fluctuates between 
$\frac{1}{(N-1)^2 +1 }$ and $\frac{1}{(N-1)^2}$ 
for $\lambda\in [0,1]$. Comparatively, the cases of 
$\lambda = 0.8$ (\textcolor{green}{\small$\vartriangle$}) 
and 0.9 (\textcolor{red}{\small$\triangledown$}) align closely 
with the $\lambda = 0$ scenario (the black line), 
exhibiting a remarkable match. Notably, 
they attain the Heisenberg scaling. 
For comparison, we show the standard quantum limit SQL = $N^{-1}$
and the Heisenberg limit HL = $N^{-2}$. 
This result represents an advanced approach in 
leveraging graph states for 
robust sensing in noisy environments,
indicating a transition from the SQL 
to the HL within a periodic range of the Larmor frequency.

\begin{figure}[t!]
    \centering
    \includegraphics[width=8.6cm]{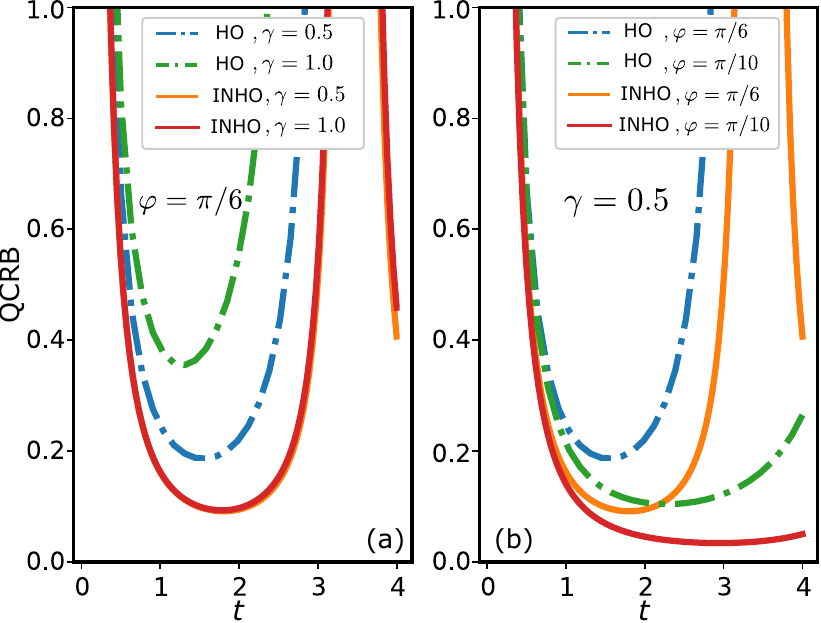}
    \caption{
    \textbf{Quantum Cram{\'e}r-Rao bound in a multiphase estimation.}
    The plot of QCRB as a function of sensing time 
    for time-homogeneous (HO) and time-inhomogeneous (INHO)
    noises.
    Here we fixed the Larmor frequencies $\phi_x = 
    \phi_y = \phi_z
    \equiv \varphi = \pi/6$ in (a)
    and fixed $\gamma = 0.5$ in (b).
    The QCRB initially decreases, reaches a minimum, 
    and then increases with increasing sensing time.
    }
    \label{fig:3}
\end{figure}

%%%%%%%%%%%%%%%%%
\subsection{Multiple phases estimation}
We consider the estimation of 
Larmor frequencies 
as $\bm \phi = (\phi_x, \phi_y, \phi_z)$
with the coupling Hamiltonian being
given in Eq.~\eqref{eq:Ht}.
The unitary evolution yields
\begin{align}\label{eq:uni_3} 
\mathcal U(\bm\phi) = \exp\big(-it\bm\phi \cdot \bm J\big). 
\end{align}

We consider the Ornstein-Uhlenbeck noise model, 
originating from the stochastic 
fluctuations of the external magnetic field 
\cite{PhysRev.36.823}.
The noise is characterized by Kraus operators 
\cite{YU2010676}
\begin{align}\label{eq:tdephasing}
\mathcal K_0(t) = {\rm diag}\big(1, \sqrt{1-q(t)}\big),
\mathcal K_1(t) = {\rm diag}\big(0, \sqrt{q(t)}\big),
\end{align}
where $q(t) = 1-e^{-f(t)}$ 
and $f(t) = \gamma[t+\tau_c (e^{-t/\tau_c}-1)]$. 
Here, $\tau_c$ is the memory time of the environment.
In the limit of
time-homogeneous
 behavior 
 or white noise limit
($\tau_c\to 0$), we have $f(t) = \gamma t$, 
which corresponds to the previous dephasing case. 
In the
time-inhomogeneous case, the time
$\tau_c$ is large, and thus $t/\tau_c\ll 1$ 
(short-time limit).
In this case, the expression becomes 
$f(t) = \frac{\gamma t^2}{2\tau_c}$.
The function $q(t)$ is defined as
\begin{align}\label{eq:timefunc}
    q(t) = \begin{cases} 1-\exp(- \gamma t) \;\;\;\; &\textrm{time-homogeneous,} \\ 
    1-\exp(-\frac{\gamma t^2}{2\tau_c}) \;\;\; &\textrm{time-inhomogeneous.} \end{cases}
\end{align}
In the numerical simulation, $\tau_c$ is fixed at $\tau_c = 20$ 
for the time-inhomogeneous case.

Similar as the single phase case, 
we first calculate $\rho(\bm\phi) = \mathcal{U}(\bm\phi)|G\rangle\langle G|\mathcal{U}^\dagger(\bm\phi)$, 
and then derive $\rho({\bm\phi,\gamma})$
by applying the Kraus operators in Eq.~\eqref{eq:tdephasing}
for all qubits.
Next, given the decomposed form as
$\rho({\bm \phi}, \gamma) 
= \sum_k \ell_k |\ell_k\rangle\langle\ell_k|$, 
the quantum Fisher information matrix (QFIM) gives
\begin{align}\label{eq:qfim}
 Q_{\alpha\beta} = 2\sum_{i,j,\ell_i+\ell_j\ne 0} 
 \dfrac{\langle\ell_i| 
 \partial_{\phi_\alpha}\rho(\bm\phi, \gamma) 
 |\ell_j\rangle \langle \ell_j|
 \partial_{\phi_\beta}\rho(\bm\phi, \gamma)|\ell_i\rangle}
 {\ell_i+\ell_j},
\end{align}
and the QCRB in the multiphase case is given by
$\mathsf{C_Q} = {\rm Tr} [Q^{-1}]$.
See detailed calculations in %App.~\ref{appC}.
the Methods section.

\begin{figure}[t!]
    \centering
    \includegraphics[width=8.6cm]{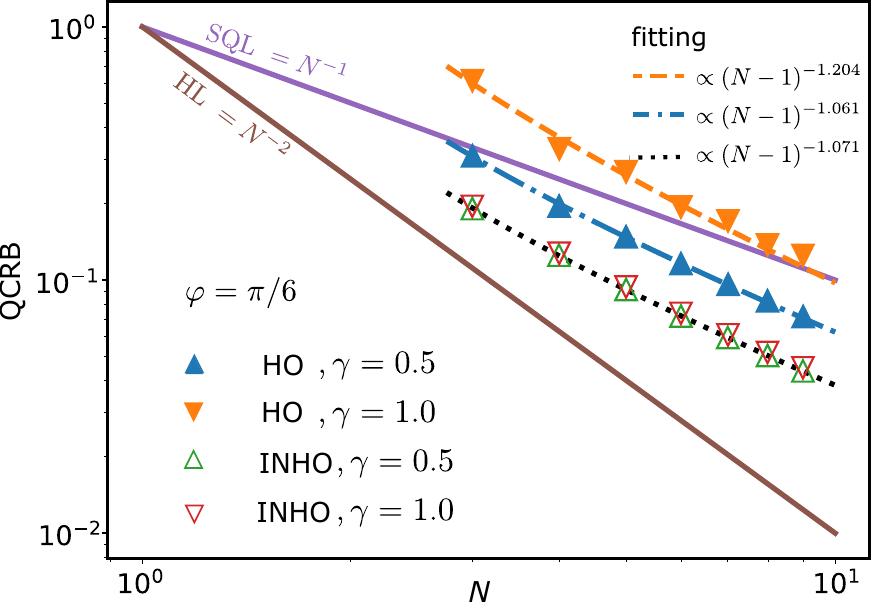}
    \caption{
    \textbf{Quantum Cram{\'e}r-Rao bound in a multiphase estimation.}
    The plot of QCRB vs $N$ at fixed Larmor frequencies 
    $\phi_x = \phi_y = \phi_z 
    \equiv \varphi = \pi/6$
    for two cases of 
    time-homogeneous and time-inhomogeneous noises.
    }
    \label{fig:4}
\end{figure}

Figure \ref{fig:3} illustrates the QCRB 
concerning  
time-homogeneous and time-inhomogeneous
noises 
as a function of time $t$. 
Consistent with findings reported in 
\cite{PhysRevA.102.022602,Le2023}, 
a pivotal insight surfaces: an optimal sensing time 
emerges leads to minimized CRBs across 
the examined scenarios. For 
time-homogeneous (HO) noise, 
the optimal sensing time tends to be shorter, 
whereas for 
time-inhomogeneous (INHO) noise, 
an extended sensing time is favored. Remarkably, 
the presence of 
time-inhomogeneous
dephasing 
yields lower metrological bounds compared to 
the
time-homogeneous counterpart.

In Figure \ref{fig:4}, we can observe the minimum 
QCRB for various values 
of $N$. The results demonstrate that as $N$ increases, 
time-inhomogeneous
noise consistently outperforms 
the SQL across all levels 
of noise (represented by open triangles). 
The relationship follows a fitted function 
that scales as $\propto (N-1)^{-1.071}$. 
Similarly, in the case of time-homogeneous noise, 
the bounds tend to surpass the SQL as $N$ grows larger.
This is the first instance we observe exceeding 
the SQL under
time-homogeneous noise.

\begin{figure}[t!]
    \centering
    \includegraphics[width=8.6cm]{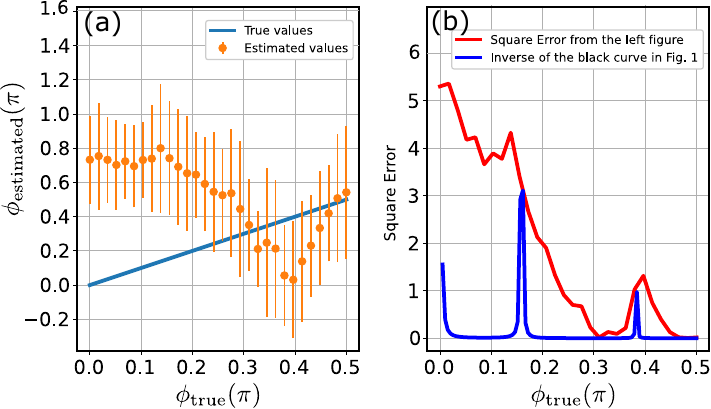}
    \caption{\textbf{Bayesian inference.}
    (a) A comparison between true and estimated values. 
    The true values are represented by the blue line where 
    $\phi_{\rm true} = \phi_{\rm estimated}$. 
    The estimated values are indicated by orange dots 
    along with error bars, derived from 100 repeated 
    experiments using Bayesian inference.
    (b) Plot of squared errors as a function of $\phi$ 
    and their comparison with the inverse QFI extracted 
    from the black curve in Fig.~\ref{fig:1}. 
    These results are presented for $\lambda = 1$.
    }
    \label{fig:5}
\end{figure}

\section{Discussion}\label{seciv}
We discuss the Bayesian inference in estimating the single Larmor frequency. 
In this approach, the process begins with defining a likelihood function 
that expresses the probability of the observed data set $\{D\}$ 
concerning the parameter of interest $\phi$. 
In our scenario, the likelihood function is calculated 
as the product of probabilities while measuring the final state
$\rho(\phi,\gamma)$ in Eq.~\eqref{eq:rho1} 
with the computational bases $\{|k\rangle\}$ as
\begin{align}\label{eq:like}
P(D|\phi) = \prod_{k=1}^{2^N} {\rm Tr} 
\Big[\rho(\phi,\gamma) |k\rangle \langle k|\Big].
\end{align}
The Bayesian theorem is then applied to compute the posterior distribution, 
which describes the uncertainty associated with the parameter as
\begin{align}\label{eq:post}
P(\phi|D) = \dfrac{P(D|\phi)}{\int P(D|\phi)d\phi}.
\end{align}
Finally, the estimated phase is given by
\begin{align}\label{eq:est}
\phi_{\rm estimated} = \int\phi P(\phi|D)d\phi.
\end{align}

Practically, sampling techniques such as Markov Chain Monte Carlo (MCMC) 
or Nested Sampling (NS) are used to generate samples 
from the posterior distribution \cite{rinaldi2023parameter}. 
These samples are then used for estimates, 
typically as the posterior mean or credible intervals. 
These estimates convey not only the point estimate 
but also the related uncertainty.

To illustrate, we focus on the case where $\lambda = 1$ 
and proceed to theoretically derive the likelihood function 
\eqref{eq:like}
using $\rho_N(\phi)$ from Eq.\eqref{eq:frhtApp}. 
It gives
\begin{widetext}
\begin{align}\label{eq:pdphimain}
\notag P(D|\phi) &= \prod_{k=1}^{2^N} {\rm Tr} \Big[\rho_N(\phi) |k\rangle \langle k|\Big]\\
&= \frac{1}{2^{N \cdot (2^N) }}  \sum_{k=0}^{2^{N-1}} 
\binom{2^{N-1}}{k} (-1)^k \sin^{2k} (\phi)\sin^{2k} [(N-1)\phi ],
\end{align}
Then $P(\phi|D)$ yeilds
\begin{align}\label{eq:ppdamain}
P(\phi|D) = \frac{1}{2^{N \cdot (2^N) }}  \sum_{k=0}^{2^{N-1}} \binom{2^{N-1}}{k} (-1)^k \int \sin^{2k} (\phi)\sin^{2k} [(N-1)\phi ] d \phi.
\end{align}
\end{widetext}
Finally, we obtain
\begin{align}\label{eq:estmain}
\phi_{\rm estimated} = \int\phi P(\phi|D)d\phi.
\end{align}
The detailed calculation for obtaining the estimated 
$\phi_{\rm estimated}$ is provided in %App.~\ref{appE}. 
the Methods section.
We present our findings in Fig.~\ref{fig:5}. 
In (a), a comparison is made between the true values 
and the estimated values. The true values are represented 
by the blue line when $\phi_{\rm true} = 
\phi_{\rm estimated}$. The average estimated values 
are indicated by orange dots with error bars, 
obtained through the Bayesian inference method from 
$M = 100$ repeated experiments in a quantum circuit. 
In (b), we plot the squared error $\Delta^2\phi$ as a function of 
$\phi$ and compare it with the inverse QFI 
extracted from the black curve in Fig.~\ref{fig:1} 
after re-scaling, i.e., 
$2.5\times\dfrac{1}{MQ}$. 
It indicates the QCRB relation
as $\Delta^2\phi \ge \dfrac{1}{MQ}$.

\section{Methods}
\subsection{Qubits model}\label{appA}
\begin{figure}[t]
    \centering
    \includegraphics[width=8.6cm]{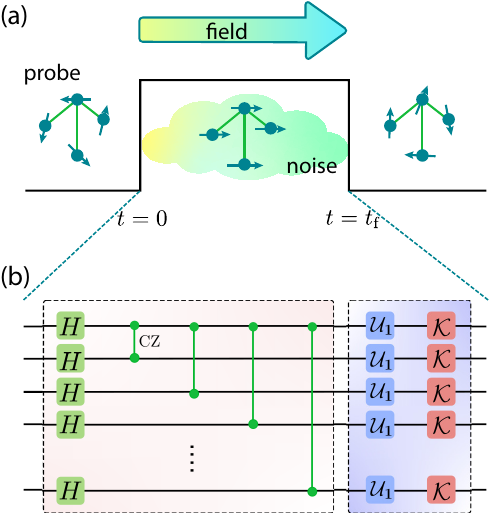}
    \caption{
    (a): Quantum magnetometry scheme with noise model. 
    The probe is an ensemble of spins prepared in a graph state. 
    The probe interacts with an external filed during the time $t = 0$ to $t = t_{\rm f}$.
    The noise also appears during the interaction time.
    (b) The quantum circuit designed for quantum magnetometry.
    The first block is a star configuration, 
    applying Hadamard gates to $|0\rangle$ qubits 
    to transform them into $|+\rangle$ states 
    and connecting them as a star, 
    with the first qubit at the center 
    linked to the surrounding qubits via CZ gates.
    The second block involves phase and noise encoding 
    using $\mathcal U_1(\bm\phi)$, followed by Kraus operators 
    $\mathcal K = (\mathcal K_0, \mathcal K_1)$ apply to all qubits.}
    \label{fig:1a}
\end{figure}

We introduce a measurement model using qubits system
as shown in Fig.~\eqref{fig:1a}.
In metrology schemes, we measure a system by using a probe that couples to it. After the interaction, we measure the probe to estimate the system's information. In our model, the system is the magnetic field, and the probe is an ensemble of spins.
When considering noise, it couples to the probe, and this coupling is different from the system-probe interaction. In our case, we consider dephasing and  Ornstein-Uhlenbeck noise models. Dephasing is often referred to as the spin-spin relaxation process. It affects the relative phase in the probe’s basis, which can be represented by the Pauli operator $\sigma_z$ .  
The noise occurs during the interaction process between the system and the probe. Initially, at $t = 0$, we turn on the coupling between the probe and the system. At time $t = t_{\rm f}$ , we turn off the interaction and measure the probe. Assuming the preparation and measurement times are very short, the probe does not evolve before time $t = 0$ and after $t = t_{\rm f}$. Thus, noise only affects the probe during the interaction time. 
The measurement scheme is given in Fig.~\eqref{fig:1a}a.

For the interaction, we first rewrite Eq.~\eqref{eq:uni_3} as
\begin{align}\label{eq:uni_3App} 
\notag \mathcal U(\bm\phi) &= \exp\Big[-it(\phi_xJ_x + \phi_yJ_y + \phi_zJ_z)\Big]\\
\notag & = \exp\Big[-it\sum_{k=1}^N \big(\frac{\phi_x}{2}\sigma_x^{(k)} 
+ \frac{\phi_y}{2}\sigma_y^{(k)} 
+ \frac{\phi_z}{2}\sigma_z^{(k)}\big) \Big] \\
&=\prod_{k=1}^N \exp\Big[-it(\frac{\phi_x}{2}\sigma_x^{(k)}
+\frac{\phi_y}{2}\sigma_y^{(k)}
+\frac{\phi_z}{2}\sigma_z^{(k)} \big) \Big].
\end{align}
We set a single-qubit unitary as
\begin{align}\label{eq:uni_1App} 
\mathcal U_1(\bm\phi) = \exp\Big[-it\big(\frac{\phi_x}{2}\sigma_x
+\frac{\phi_y}{2}\sigma_y
+\frac{\phi_z}{2}\sigma_z \big) \Big],
\end{align}
and apply it to all qubits in a quantum circuit.
For a single-phase estimation, it becomes 
a rotation gate, i.e. 
$\mathcal U_1(\bm\phi) = \exp\big[-it\frac{\phi_x}{2}\sigma_x$\big].

The quantum circuit is given in Fig.~\ref{fig:1a}b. 
The first block is the graph state generation with 
star configuration, wherein a Hadamard gate 
is applied to each qubit initially set 
in the state $|0\rangle$, 
transforming them into $|+\rangle$ states. 
These qubits are connected as a star, 
with the first qubit at the center 
and connects to the surrounding qubits through CZ gates.
The second block is 
phase and noise encoding, given by applying $\mathcal U_1(\bm\phi)$
and followed by the Kraus operators 
$\mathcal K = (\mathcal K_0, \mathcal K_1)$ apply to all qubits.
The final state is used to calculate 
quantum Fisher information (QFI) for single phase estimation and quantum Fisher information matrix (QFIM) for multiphase estimation. 
From these, we derive the corresponding QCRB
\cite{Ho2023}.

\subsection{Deriving QFI for single parameter estimation}\label{appB}
\subsubsection{For star graph state}
We first calculate the QFI 
with the initial star graph state 
in the case of without noise.
We express the QFI in terms of its generator as
\begin{align}\label{eq:qfia}
    Q = 4\langle \Delta \mathcal H^2_\phi \rangle
    = 4\Big[\langle G| \mathcal H^2_\phi |G \rangle 
    - (\langle G| \mathcal H_\phi |G \rangle)^2\Big],
\end{align}
where $\mathcal H_\phi = 
i\mathcal U^{\dagger}(\phi) 
\partial_\phi \mathcal U(\phi)$.
For single-phase estimation, 
it gives $\mathcal H_\phi = J_x = \frac{1}{2}
\sum_{k = 1}^N \sigma_x^{(k)}$.

To calculate Eq.~\eqref{eq:qfia}, 
we first expand the graph state to 
a star configuration, 
with one central qubit (qubit 1)
connects to the remaining surrounding $N-1$ qubits:
\begin{align}\label{eq:wpa}
    \notag |G \rangle &=  \prod_{k=2}^N {\rm CZ}_{1,k}|+\rangle^{\otimes N}\\
    \notag &= \Big({\rm CZ}_{1,N} {\rm CZ}_{1,N-1}\cdots {\rm CZ}_{1,2}\Big) |+\rangle^{\otimes N}\\
    & = \dfrac{1}{\sqrt{2}}\Big(|0\rangle|+\rangle^{\otimes (N-1)} 
        +|1\rangle|-\rangle^{\otimes (N-1)}\Big).
\end{align}
Next, we compute $\mathcal H_\phi |G\rangle$:
%\begin{widetext}
\begin{align}\label{eq:Hwpa}
    \notag \mathcal H_\phi|G \rangle &=  \Big[\dfrac{1}{2}\sum_{k=1}^N\sigma_x^{(k)}\Big]|G\rangle\\
    %\notag &= \dfrac{1}{2}\Big[\dfrac{1}{\sqrt{2}}\Big(|1\rangle|+\rangle^{\otimes (N-1)} 
    %                                                        +|0\rangle|-\rangle^{\otimes (N-1)}\Big) + 
    %            \dfrac{1}{2}\Big[\dfrac{1}{\sqrt{2}}\Big(|0\rangle|+\rangle^{\otimes (N-1)} 
    %                                                        -|1\rangle|-\rangle^{\otimes (N-1)}\Big)
    %            +\cdots+ \dfrac{1}{2}\Big[\dfrac{1}{\sqrt{2}}\Big(|0\rangle|+\rangle^{\otimes (N-1)} 
    %                                                        -|1\rangle|-\rangle^{\otimes (N-1)}\Big)\Big]\\
    \notag & = \dfrac{1}{2\sqrt{2}}\Big[\Big(|1\rangle|+\rangle^{\otimes (N-1)} 
                                                            +|0\rangle|-\rangle^{\otimes (N-1)}\Big) + \\
                & \hspace{1cm} (N-1)\Big(|0\rangle|+\rangle^{\otimes (N-1)} 
                                                            -|1\rangle|-\rangle^{\otimes (N-1)}\Big)
    \Big].
\end{align}
%\end{widetext}
%
Using Eq.~\eqref{eq:Hwpa}, the first term in Eq.~\eqref{eq:qfia} gives:
\begin{align}\label{eq:qfia1}
     \langle G| \mathcal H^2_\phi |G \rangle 
    = \langle G| \mathcal H_\phi \mathcal H_\phi |G \rangle
    =\dfrac{1}{4}\Big(1+(N-1)^2\Big).
\end{align}
Using Eqs.~(\ref{eq:wpa},\ref{eq:Hwpa}), 
the second term in Eq.~\eqref{eq:qfia} yields
\begin{align}\label{eq:qfia2}
     \langle G| \mathcal H_\phi |G \rangle^2 = 0.
\end{align}
As a result, the QFI in \eqref{eq:qfia} gives
\begin{align}\label{eq:qfiaf}
    Q = 1 + (N-1)^2,
\end{align}
and the corresponding QCRB is 
$\mathsf{C_Q} = \frac{1}{Q} = \frac{1}{1 + (N-1)^2}$.

Now, we calculate the QFI under dephasing noise. 
We first recast the unitary Eq.~\eqref{eq:uni} as 
\begin{align}\label{eq:uniApp} 
\notag \mathcal U(\phi) &= \Big[\exp\big(-i\frac{\phi}{2}
    \sigma_x\big)\Big]^{\otimes N}\\
    & = \Big[
    \cos(\phi') I 
- i \sin(\phi')\sigma_x
    \Big]^{\otimes N},
\end{align}
where we fixed $t = 1$ and set $\phi' = \phi/2$. 
The initial graph state $|G\rangle$ evolves to

\begin{widetext}
\begin{align}\label{eq:evlstarApp}
\notag |G(\phi) \rangle &= 
    \Big[
    \cos(\phi') I 
    - i \sin(\phi')\sigma_x
    \Big]^{\otimes N}\cdot
    \dfrac{1}{\sqrt{2}}\Big(|0\rangle|+\rangle^{\otimes (N-1)} 
        +|1\rangle|-\rangle^{\otimes (N-1)}\Big)
    \\
\notag&= \dfrac{1}{\sqrt{2}} 
    \Big\{\Big(\cos(\phi')|0\rangle
    - i \sin(\phi')|1\rangle\Big)
    \otimes 
    e^{-i\phi'(N-1)}|+\rangle^{\otimes (N-1)} + \\
\notag&\hspace{4cm}    \Big(\cos(\phi')|1\rangle
    - i \sin(\phi')|0\rangle\Big)
    \otimes 
    e^{i\phi'(N-1)}|-\rangle^{\otimes (N-1)}\Big\}\\
    &= \dfrac{1}{\sqrt{2}}
    \Bigg\{
    e^{-i\phi'(N-1)}
    \begin{pmatrix}\cos\phi' \\ -i \sin\phi'\end{pmatrix}
    \otimes|+\rangle^{\otimes (N-1)}
    + e^{i\phi'(N-1)}
    \begin{pmatrix}-i \sin\phi' \\ \cos\phi' \end{pmatrix}
    \otimes|-\rangle^{\otimes (N-1)}
    \Bigg\}.
\end{align}
\end{widetext}
%where $|R(L)\rangle = \frac{1}{\sqrt{2}}(|0\rangle \pm i|1\rangle)$
%are the two orthogonal $y$-basis states.
The single-qubit dephasing is represented by 
Kraus operators $\mathcal K_0 = {\rm diag}(1, \sqrt{1-\lambda}),
\mathcal K_1 = {\rm diag}(0, \sqrt{\lambda})$. 
They act on a qubit $i$ as $\mathcal K_0^{(i)} = 
I\otimes\cdots \mathcal K_0\cdots\otimes I$
and $\mathcal K_1^{(i)} = 
I\otimes\cdots \mathcal K_1\cdots\otimes I$.
Under dephasing, the quantum state 
\eqref{eq:kraus} explicitly gives
\begin{align}\label{eq:fullApp}
\rho_{k}(\phi) = 
\mathcal K_0^{(k)}\rho_{(k-1)}(\phi)\big[\mathcal K_0^{(k)}\big]^\dagger+
\mathcal K_1^{(k)}\rho_{(k-1)}(\phi)\big[\mathcal K_1^{(k)}\big]^\dagger,
\end{align}
for $k = 1,\cdots, N$ 
with $\rho_{0}(\phi) = |G(\phi) \rangle \langle G(\phi)|$.

%\begin{align}\label{eq:krausApp}
% \rho_{\rm f} &= 
%    \begin{pmatrix}
%        1&0\\0&\sqrt{1-\lambda}
%    \end{pmatrix}^{\otimes N}
%    |G(\phi) \rangle \langle G(\phi)|
%    \begin{pmatrix}
%        1&0\\0&\sqrt{1-\lambda}
%    \end{pmatrix}^{\otimes N}
%+\begin{pmatrix}
%        0&0\\0&\sqrt{\lambda}
%    \end{pmatrix}^{\otimes N}
%    |G(\phi) \rangle \langle G(\phi)|
%    \begin{pmatrix}
%        0&0\\0&\sqrt{\lambda}
%    \end{pmatrix}^{\otimes N}.
%\end{align}

To simplify the calculation, we focus on 
the maximum noise probability, $\lambda = 1$.
We first calculate 
%\begin{widetext}
\begin{align}\label{eq:ftApp}
\notag \rho_1(\phi) &= \mathcal K_0^{(1)}
    |G(\phi) \rangle \langle G(\phi)|
    \mathcal K_0^{(1)}+
    \mathcal K_1^{(1)}
    |G(\phi) \rangle \langle G(\phi)|
    \mathcal K_1^{(1)}\\
\notag& = \dfrac{1}{2}\Bigg\{
    \begin{pmatrix}
        \cos^2\phi'&0\\0&\sin^2\phi'
    \end{pmatrix}\otimes \big(|+\rangle\langle+|\big)^{\otimes (N-1)}\\
\notag&\hspace{1cm}+e^{-2i\phi'(N-1)}\dfrac{i}{2}\sin(2\phi')\sigma_z\otimes \big(|+\rangle\langle-|\big)^{\otimes (N-1)}\\
\notag&\hspace{1cm}+e^{2i\phi'(N-1)}\dfrac{-i}{2}\sin(2\phi')\sigma_z\otimes \big(|-\rangle\langle+|\big)^{\otimes (N-1)}\\
      &\hspace{1cm} +\begin{pmatrix}
        \sin^2\phi'&0\\0&\cos^2\phi'
    \end{pmatrix}\otimes \big(|-\rangle\langle-|\big)^{\otimes (N-1)}
    \Bigg\},
\end{align}
%\end{widetext}

Continuously we calculate $\rho_2(\phi)$ as:
\begin{align}\label{eq:ftApp1}
\notag \rho_2(\phi) &= \mathcal K_0^{(2)} \rho_1(\phi) K_0^{(2)}+
    \mathcal K_1^{(2)} \rho_1(\phi) K_1^{(2)} \\
\notag&= \dfrac{1}{4}\Bigg\{
    \begin{pmatrix}
        \cos^2\phi'&0\\0&\sin^2\phi'
    \end{pmatrix}\otimes I \otimes \big(|+\rangle\langle+|\big)^{\otimes (N-2)}\\
\notag&\hspace{1cm}+e^{-2i\phi'(N-1)}\dfrac{i}{2}\sin(2\phi')\sigma_z\otimes  \sigma_z\otimes \big(|+\rangle\langle-|\big)^{\otimes (N-2)}\\
\notag&\hspace{1cm}+ e^{2i\phi'(N-1)}\dfrac{-i}{2}\sin(2\phi')\sigma_z\otimes \sigma_z\otimes \big(|-\rangle\langle+|\big)^{\otimes (N-2)}\\
      &\hspace{1cm} +\begin{pmatrix}
        \sin^2\phi'&0\\0&\cos^2\phi'
    \end{pmatrix}\otimes I \otimes \big(|-\rangle\langle-|\big)^{\otimes (N-2)}
    \Bigg\},
\end{align}
and so on.
Finally, we get
\begin{align}\label{eq:frhtApp}
    \rho_N(\phi)=\dfrac{1}{2^N}
    \Big(I^{\otimes N}+\sin(\phi)\sin[(N-1)\phi]\sigma_z^{\otimes N}
    \Big).
\end{align}

To calculate the QFI, we use Eq.~\eqref{eq:qfi}
in the main text with $\rho_N({\phi}) 
= \sum_k \ell_k |\ell_k\rangle\langle\ell_k|$, 
and get
\begin{align}\label{eq:qfiApp}
 Q = 2\sum_{i,j,\ell_i+\ell_j\ne 0} 
 \dfrac{\big|\langle\ell_i| 
 \partial_\phi\rho_{N}(\phi) 
 |\ell_j\rangle\big|^2}{\ell_i+\ell_j} 
 = (N-1)^2,
\end{align}
where we first calculated $\partial_\phi\rho_{N}(\phi)$
from Eq.~\eqref{eq:frhtApp}, and then used $\phi = \pi/2$.

\subsubsection{For GHZ state}
We consider the case where
the initial probe state is a GHZ state.
For the generator 
$\mathcal H_\phi = J_x = \frac{1}{2}
\sum_{k = 1}^N \sigma_x^{(k)}$, 
the defined GHZ state in Eq.~\eqref{eq:ghz} 
explicitly gives:
\begin{align}\label{eq:ghzApp}
    |\psi_{\rm GHZ}\rangle=\dfrac{1}{\sqrt{2}}
    \Big(|+\rangle^{\otimes N} 
    + |-\rangle^{\otimes N}\Big),
\end{align}
where $|\pm\rangle^{\otimes N}$
are eigenstates of $\mathcal H_\phi$ corresponds 
to the maximum and minimum eigenvalues.
This state can be prepared from
a star-graph state by adding a Hadamard gate
into the first qubits of Eq.~\eqref{eq:wpa}.

We first compute two terms 
$\mathcal{H}_\phi|\psi_{\rm GHZ}\rangle$ 
and $\mathcal{H}^{2}_\phi|\psi_{\rm GHZ}\rangle$, 
where
\begin{align}
   \notag \mathcal{H}_\phi|\psi_{\rm GHZ}\rangle 
    &= \Big[\dfrac{1}{2}\sum_{k=1}^N\sigma_x^{(k)}\Big] 
    \frac{1}{\sqrt{2}} \Big( |+\rangle^{\otimes N} 
    + |-\rangle^{\otimes N} \Big) \\
    &= \dfrac{N}{2 \sqrt{2}}
    \Big(|+\rangle^{\otimes N} - |-\rangle^{\otimes N}\Big),
\end{align}
and 
\begin{align}
    \mathcal{H}^{2}_\phi|\psi_{\rm GHZ}\rangle 
    &=  \dfrac{N^2}{4}|\psi_{\rm GHZ}\rangle.
\end{align}
Then, we obtain $\langle\psi_{\rm GHZ}|\mathcal{H}_\phi|\psi_{\rm GHZ}\rangle
= 0$ and $\langle\psi_{\rm GHZ}|\mathcal{H}_\phi^2|\psi_{\rm GHZ}\rangle
= N^2/4$.
Finally, the QFIM yields 
\begin{align}\label{eq:QghzApp}
    \notag Q_{\rm GHZ} &= 4\Big[\langle \psi_{\rm GHZ}| \mathcal{H}_\phi^2 |\psi_{\rm GHZ} \rangle 
    - \big(\langle \psi_{\rm GHZ}| \mathcal{H}_\phi |\psi_{\rm GHZ} \rangle\big)^2\Big]\\ 
    & = N^2.
\end{align}
Similarly, for noisy cases, we have 
$Q_{\rm GHZ} = N^2$.

\subsection{Deriving QFIM for multiparameter estimation}\label{appC}
The QFIM for a pure star graph state is given by 
\begin{align}\label{eq:QFIM}
Q_{\alpha\beta} = 
\dfrac{1}{2}
\langle G({\bm\phi})|\bigl(L_\alpha L_\beta
+L_\beta L_\alpha\bigr)|G({\bm\phi})\rangle,
%4{\rm Re}\bigl[\langle\partial_{\phi_\alpha}
%\psi_{\bm\phi}|\partial_{\phi_\beta}\psi_{\bm\phi}\rangle
%-\langle\partial_{\phi_\alpha}\psi_{\bm\phi}|\psi_{\bm\phi}\rangle
%\langle\psi_{\bm\phi}|\partial_{\phi_\beta}\psi_{\bm\phi}\rangle\bigr].
\end{align}
where $L$ is given in the symmetric logarithmic derivative 
(SLD) as 
\begin{align}\label{eq:L}
L_{\alpha} = 2\bigl(|\partial_{\phi_\alpha}G({\bm\phi})\rangle\langle G({\bm\phi})|
+|G({\bm\phi})\rangle\langle\partial_{\phi_\alpha}G({\bm\phi})|\bigr).
\end{align}
%for an arbitrary pure state $|\psi_{\bm\phi}\rangle$. 
For concreteness, we first derive 
%\begin{align}\label{eq:der_psi_phi}
%|\partial_{\phi_\alpha}G({\bm\phi})\rangle 
%= \partial_{\phi_\alpha}\mathcal U(\bm\phi)|G\rangle
%=-i\mathcal U(\bm\phi)\mathcal A_\alpha|G\rangle,
%\end{align}
%%%%%%%%%%%
\begin{align}\label{eq:der_psi_phi_temp}
\notag |\partial_{\phi_\alpha}G({\bm\phi})\rangle 
&= \partial_{\phi_\alpha}\mathcal U(\bm\phi)|G\rangle\\
\notag &= \partial_{\phi_\alpha}
e^{-it\mathcal H}|G\rangle\\
\notag &= -i\int_0^t due^{-i(1-u)\mathcal H}
[\partial_{\phi_\alpha}\mathcal H]e^{-iu\mathcal H}|G\rangle\\
\notag &= -ie^{-i\mathcal H}\int_0^t due^{iu\mathcal H}
J_\alpha e^{-iu\mathcal H}|G\rangle\\
&=-i\mathcal U(\bm\phi)\mathcal A_\alpha|G\rangle,
\end{align}
where $\mathcal H$ is given in Eq.~\eqref{eq:Ht},
and 
\begin{align}\label{eq:Aalpha}
\mathcal A_\alpha = \int_0^t du\ e^{iu \mathcal H}J_\alpha
e^{-iu \mathcal H},
\end{align}
is a Hermitian operator \cite{10.1063/1.1705306,PhysRevA.102.022602}.

Then, the SLD \eqref{eq:L} and QFIM \eqref{eq:QFIM} are explicitly given as
\begin{align}
L_{\alpha}&=
2i\mathcal U(\bm{\phi})\bigl[|G\rangle\langle G|,
\mathcal A_{\alpha}\bigr]\mathcal U^\dagger(\bm\phi), \label{eq:SLD:con}\\
Q_{\alpha\beta} &= 
4{\rm Re}\bigl[\langle G|\mathcal A_\alpha \mathcal A_\beta|G\rangle
-\langle G|\mathcal A_\alpha|G\rangle
\langle G|A_\beta|G\rangle\bigr].\label{eq:QFIM:con}
\end{align}
In quantum circuits, the QFIM can be calculated using
a stochastic method \cite{Ho2023}.

For a general mixed state, such as a star graph under noise,
i.e., $\rho({\bm \phi}, \gamma) 
= \sum_k \ell_k |\ell_k\rangle\langle\ell_k|$, 
the QFIM gives
\begin{align}\label{eq:qfimApp}
 Q_{\alpha\beta} = 2\sum_{i,j,\ell_i+\ell_j\ne 0} 
 \dfrac{\langle\ell_i| 
 \partial_{\phi_\alpha}\rho(\bm\phi,\gamma) 
 |\ell_j\rangle \langle \ell_j|
 \partial_{\phi_\beta}\rho(\bm\phi,\gamma)|\ell_i\rangle}
 {\ell_i+\ell_j}.
\end{align}
The QCRB in this case is given by
$\mathsf{C_Q} = {\rm Tr} [Q^{-1}]$.

\begin{figure}[t]
    \centering
    \includegraphics[width=8.6cm]{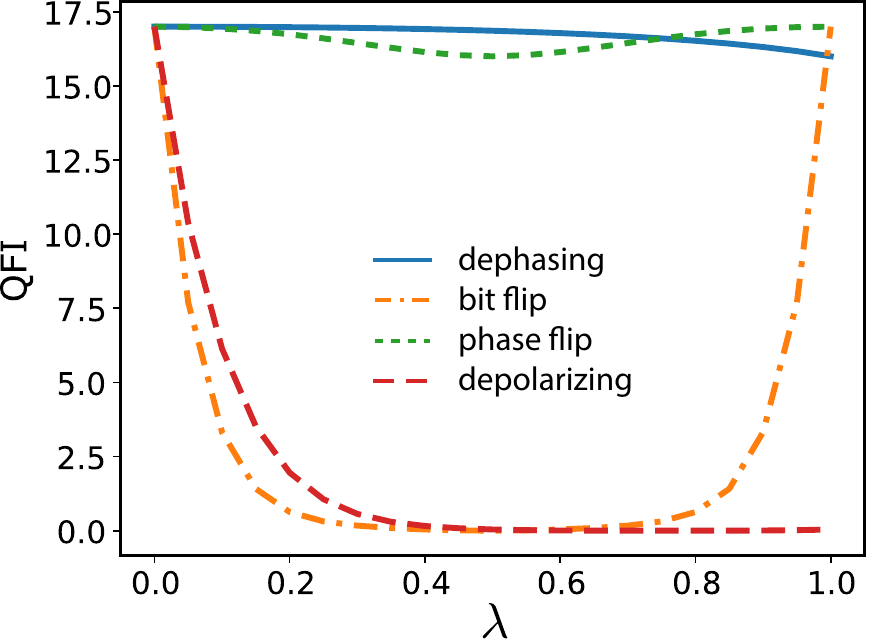}
    \caption{QFI under different noises.}
    \label{fig:2a}
\end{figure}

\subsection{Noisy QFI of graph states}\label{appD}
In this section, we examine how different types of noise 
impact the QFI. In addition to dephasing, 
we also encounter bit flip, phase flip, 
and depolarizing. The relevant Kraus 
operators are as follows:

$
%\begin{align}\notag
K_0 = \sqrt{1-\gamma}\begin{pmatrix}1&0\\0&1\end{pmatrix};
K_1 = \sqrt{\gamma}\begin{pmatrix}0&1\\1&0\end{pmatrix},
\text{ for bit flip},
%\end{align}
$
$
%\begin{align}\notag
K_0 = \sqrt{1-\gamma}\begin{pmatrix}1&0\\0&1\end{pmatrix};
K_1 = \sqrt{\gamma}\begin{pmatrix}1&0\\0&-1\end{pmatrix},
\text{ for phase flip},
%\end{align}
$
$
%\begin{align}\notag
    K_0 = \sqrt{1-\gamma}\begin{pmatrix}1&0\\0&1\end{pmatrix};
    K_1 = \sqrt{\gamma/3}\begin{pmatrix}0&1\\1&0\end{pmatrix};
    K_2 = \sqrt{\gamma/3}\begin{pmatrix}0&-i\\i&0\end{pmatrix};
    K_3 = \sqrt{\gamma/3}\begin{pmatrix}1&0\\0&-1\end{pmatrix},
    \text{ for depolarizing},
%\end{align}
$
where $\gamma\in[0,1]$ the noise probability. 
The QFI is shown in Fig.~\ref{fig:2a},
with different noises.

\subsection{Bayes Inference}\label{appE}
We start from the final state
    \begin{align}
    \rho_N(\phi)=\dfrac{1}{2^N}
    \Big(I^{\otimes N}+\sin(\phi)\sin[(N-1)\phi]\sigma_z^{\otimes N}
    \Big),
\end{align}
when $\lambda = 1$.
Because $\sigma_z^{\otimes N}$ 
has the same numbers of component $+1$ and $-1$, 
the state would contain the same number of components 
$1 + \sin(\phi)\sin[(N-1)\phi] $ and 
$1 - \sin(\phi)\sin[(N-1)\phi]$ 
on the diagonal. Therefore
\begin{widetext}
\begin{align}
\notag P(D|\phi) &= \prod_{k=1}^{2^N} {\rm Tr} \Big[\rho_N(\phi) |k\rangle \langle k|\Big]\\
\notag &= \frac{1}{2^{N \cdot (2^N)}} 
\Bigg[ ( 1 + \sin(\phi)\sin[(N-1)\phi] )^{2^{(N-1)}} ( 1 - \sin(\phi)\sin[(N-1)\phi] )^{2^{(N-1)}} \Bigg] \\
\notag &= \frac{1}{2^{N \cdot (2^N) }} \Bigg[  1 - \sin^2 (\phi)\sin^2 [(N-1)\phi ] \Bigg] ^{2^{(N-1)}}\\
&= \frac{1}{2^{N \cdot (2^N) }}  \sum_{k=0}^{2^{N-1}} 
\binom{2^{N-1}}{k} (-1)^k \sin^{2k} (\phi)\sin^{2k} [(N-1)\phi ],
\end{align}
\end{widetext}
%With $\lambda_k$ is the computational basis. 
And then $P(\phi|D)$ is 
\begin{widetext}
\begin{align}\label{eq:ppda}
\notag P(\phi|D) = \int P(D|\phi)d\phi &=  \frac{1}{2^{N \cdot (2^N) }} \int \sum_{k=0}^{2^{N-1}} \binom{2^{N-1}}{k} (-1)^k \sin^{2k} (\phi)\sin^{2k} [(N-1)\phi ] d \phi \\
    &= \frac{1}{2^{N \cdot (2^N) }}  \sum_{k=0}^{2^{N-1}} \binom{2^{N-1}}{k} (-1)^k \int \sin^{2k} (\phi)\sin^{2k} [(N-1)\phi ] d \phi.
\end{align}
\end{widetext}
Finally, we obtain
\begin{align}\label{eq:estApp}
\phi_{\rm estimated} = \int\phi P(\phi|D)d\phi.
\end{align}
The integrals in Eqs. (\ref{eq:ppda}, \ref{eq:estApp})
are derived numerically.

\section{Conclusion}\label{secv}

This study focuses on precise measurement techniques in noisy quantum systems. By using graph-state resources, we developed a method that enhances measurement accuracy and resilience against noise.
Particularly, we used
graph-state resources for robust 
quantum magnetometry under noise and showed promise 
for overcoming practical measurement challenges. 
Our demonstrations highlight significant advancements 
in accurately measuring both single and 
multiple Larmor frequencies. These outcomes 
showcase a spectrum ranging from surpassing 
the standard quantum limit to achieving Heisenberg scaling, 
marking significant progress.

These advancements are crucial across quantum computing, communication, and sensing, where precise measurements are indispensable. The capability to achieve high-precision measurements despite noise underscores the reliability and efficiency of quantum technologies.
Further research in graph state-based quantum metrology could revolutionize various fields and pave the way for practical quantum technologies despite noise.
We remark that with advancements in experimental generation 
of arbitrary photonic graph states via atomic \cite{Thomas2022} sources, 
and quantum-error-correcting code 
based on graph states \cite{Vigliar2021}, this line of research will provide 
a theoretical basis for quantum-metrological 
advantages through experiments with graph states. 

\section*{Data availability}
Data are available from the corresponding authors upon reasonable request.

\section*{Code availability}
All codes used to produce the findings of this study are 
incorporated into \texttt{tqix} \cite{VIET2023108686,HO2021107902}
and available at: https://github.com/echkon/tqix-developers.

%\begin{acknowledgments}
%This work was supported by 
%JSPS KAKENHI Grant Number 23K13025.
%\end{acknowledgments}

\section*{Acknowledgements}
This work is supported by 
JSPS KAKENHI Grant Number 23K13025.

\section*{Author contributions statement}
P.T.N and T.K.L. wrote the initial code and 
implemented the numerical simulation.
L.B.H. derived the theoretical framework, 
implemented the numerical simulation, 
and analyzed the results. 
L.B.H. and H.Q.N supervised the work.
All authors discussed 
and wrote the manuscript.

\section*{Competing interests}
The author declares no competing interests.

\bibliography{refs}

\end{document}